\documentclass[seceq]{ptptex}

\usepackage{graphicx}
\usepackage{wrapft}

\title{%        %You can use \\ for explicit line-break
Glassy dynamics and nonextensive effects in the HMF model: the importance of  initial conditions\footnote{\bf{Published in Progr. Theor. Phys. Suppl. 162 (2006) 18}}%
}

\author{%       %Use \scshape  for the family name
Alessandro \textsc{Pluchino} and Andrea \textsc{Rapisarda}%
}
\inst{%     %Affiliation, neglected when [addenda] or [errata]
Dipartimento di Fisica e Astronomia and INFN Universit\'a di Catania
Via S. Sofia 64, 95123  Catania, Italy. % www.ct.infn.it/~cactus
}

\abst{%         %this abstract is neglected when [addenda] or [errata]
We review the anomalies of the HMF model and discuss the robusteness of the glassy 
features vs the initial conditions. Connections to Tsallis statistics are also addressed.
}

\begin{document}

\maketitle

\section{Introduction}

In this work we discuss  glassy dynamics and nonextensive thermodynamics 
in the context of the
so called \textit{Hamiltonian Mean Field} (HMF) model, a
system of inertial spins with long-range interaction.
This model and its generalizations 
has been intensively investigated in the last
years for its anomalous dynamical behavior
\cite{hmf0,hmf1,hmf-dif,hmf2,hmf3,alfaxy,plu1,celia1,plu2,plu3,yama,chava,erice}.
With respect to systems with short-range interactions,
the dynamics and the thermodynamics of many-body systems
of particles interacting with long-range forces,
as the HMF, are particularly rich and interesting.

The Hamiltonian Mean Field model 
is exactly solvable at equilibrium, but exhibits a series
of anomalies in the dynamics, as the presence
of quasistationary states (QSS) characterized by:
anomalous diffusion, vanishing Lyapunov exponents,
non-gaussian velocity distributions, aging and  fractal-like
phase space structure.
%Furthermore, the model is easily accessible by means
%of both molecular dynamics and Monte Carlo simulations.
Thus, it represents a very useful ``laboratory'' for exploring
metastability and  dynamical anamalies in systems with long-range
interactions.
The model can be considered as a  pedagogical model
for a large class of complex systems,
among which one can surely include self-gravitating  \cite{chava}
and glassy systems \cite{plu2}, but also systems apparently
belonging to different fields as biology or sociology. Similar features
 were recently found  also in the context of
the \textit{Kuramoto model} \cite{kuramoto}, one of the
simplest models for synchronization in biological
systems \cite{kurastab}. Moreover,
the proliferation of metastable states in the vicinity of
a critical point in the phase diagram seems to be
quite a general feature
~\cite{parisi-science}~.
\\
In this paper we focus on two different aspects of the HMF model:
its glassy-dynamics and the possible connections with the generalized
thermodynamics. We will study in particular the dependence of these features 
on the initial conditions considered. 

\section{Out-of-equilibrium dynamical anomalies}

The Hamiltonian of the HMF model is given by 
\begin{equation}
H = K + V = \sum_{i=1}^N \frac{p_i^2}{2} + \frac{1}{2N}
\sum_{i,j=1}^N [1-\cos(\theta_i - \theta_j)]~~~~.
\label{model0}
\end{equation}
\noindent This Hamiltonian  describes  classical
$XY$-spins (or  rotators) with unitary mass and infinite
range coupling, but it  can also represent particles moving on the
unit circle. In the latter case the coordinate
$\theta_i$ of the  $i-th$ particle is its position on the circle and $p_i$
is its conjugate momentum (or angular  velocity). 
The potential energy term is rescaled by $1/N$ in order
to get a finite energy density  in the thermodynamic limit $N
\rightarrow \infty$ ~\cite{hmf0,hmf1}.
Associating to each particle
the spin vector
%\begin{displaymath}
$\overrightarrow{s}_i =(\cos \theta_i, \sin \theta_i)~~$,
%\end{displaymath}
one can  introduce the
following mean-field order parameter
%\begin{equation}
$M = \frac{1}{N} |\sum_{i=1}^N \overrightarrow{s}_i |~~$,
%\label{m0}
%\end{equation}
which is the modulus of the total {\em magnetization}.
\\
At  equilibrium the exact solution predicts a
second-order phase transition. The system passes as a function of energy 
 from a low-energy condensed
(ferromagnetic) phase with magnetization  $M\ne0$, to a
high-energy one (paramagnetic), where the spins are homogeneously
oriented on the unit circle and $M=0$. The {\em caloric curve},
i.e. the dependence of the energy density $U = H/N$ on the
temperature $T$, is given by
$U = \frac{T}{2} + \frac{1}{2} \left( 1 - M^2 \right)
~~$\cite{hmf0,hmf2}~.
The critical point is found at a temperature $T_c=0.5$, which 
corresponds to the critical energy density $U_c=0.75$.
\\
At variance with the equilibrium scenario, the out-of-equilibrium
dynamics shows, just below the phase transition, several anomalies
before  equilibration. More precisely, starting from water bag initial conditions 
in the momenta, i.e. velocities uniformly distributed, 
and a fully magnetized state, i.e.
all the  $\theta_i=0$ so that M(0)=1, the results of the simulations, in a special region
of energy values (in particular for $0.68<U<U_c$) show a disagreement with
the equilibrium  prediction for a transient regime whose length
depends on the system size N. In such a regime the system remains
trapped in metastable quasi-stationary states (QSS) with vanishing magnetization
at a temperature lower then the equilibrium one and shows strong memory effects,
correlations and aging. Then it slowly relaxes towards
Boltzmann-Gibbs (BG) equilibrium. This transient QSS regime becomes stable
if one takes the infinite size limit before the infinite time
limit~\cite{hmf3} .

\subsection{Hierarchical structures and glassy dynamics}

In order to investigate the nature of this trapping phenomenon, it
is interesting to explore directly the microscopic evolution of
the QSS. A way to visualize it is by plotting the time evolution of
the \textit{Boltzmann $\mu$-space}: each particle of the
system is represented by a point in a plane characterized by the
conjugate variables $\theta_i$ and  $p_i$.
It has been shown\cite{plu1} that, during the QSS regime,
correlations, structures and clusters formation in the $\mu$-space
appear for the $M(0)=1$ initial conditions ($M1 ~ic$), but not for initial 
conditions with zero initial magnetization ($M0 ~ic$). 
In fact, while the former case is characterized by a sudden quenching from 
an high temperature state that causes a strong mixing of particles in the metastable regime, 
in the latter case no quenching is present:
both the angles and velocities distributions remain homogeneous from the beginning
and a very slow mixing of the particles can be observed.
\begin{figure} [t]
\centerline{\includegraphics[width=7.cm,height=7. cm]
{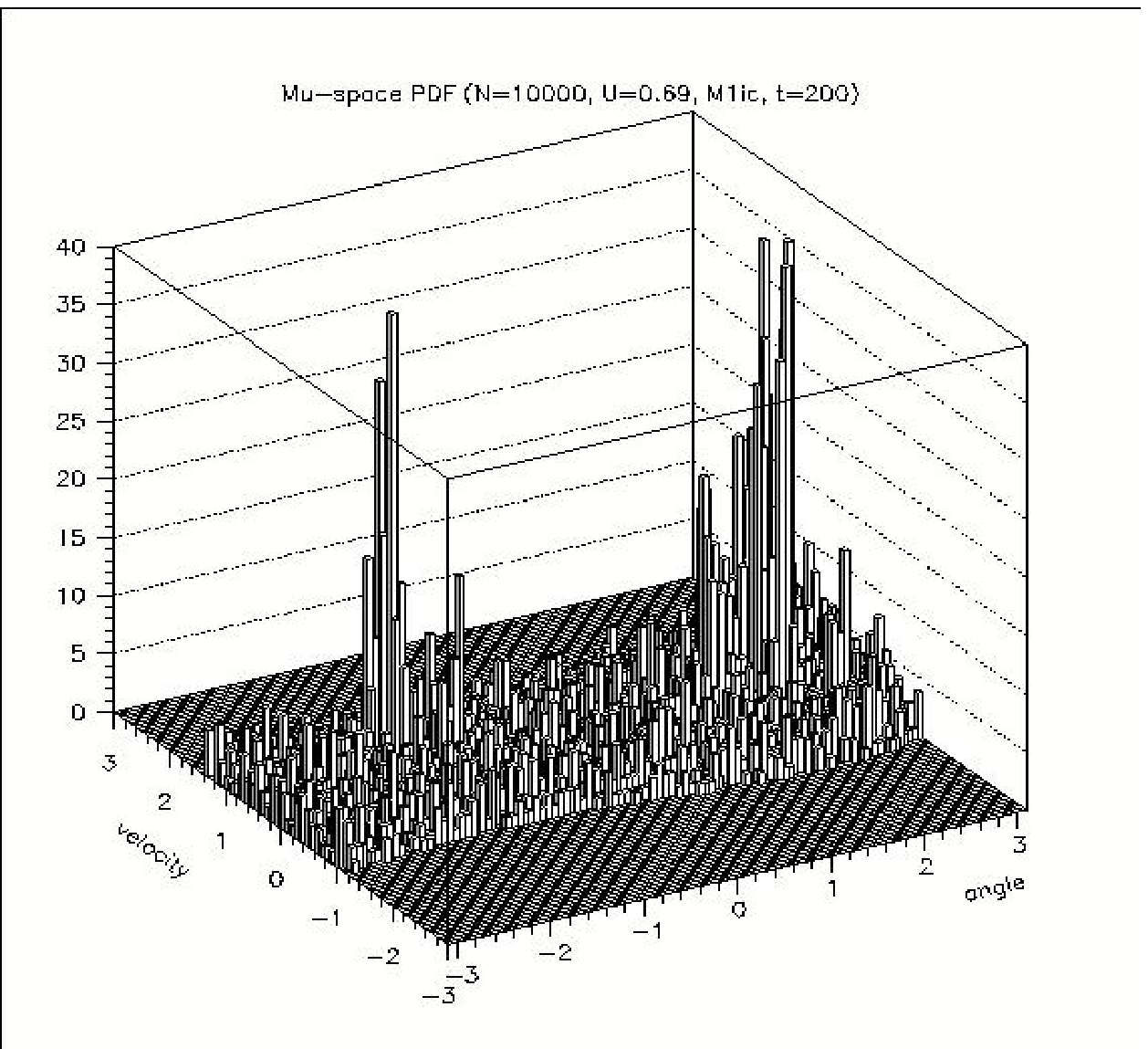}\includegraphics[width=7.cm,height=7. cm]{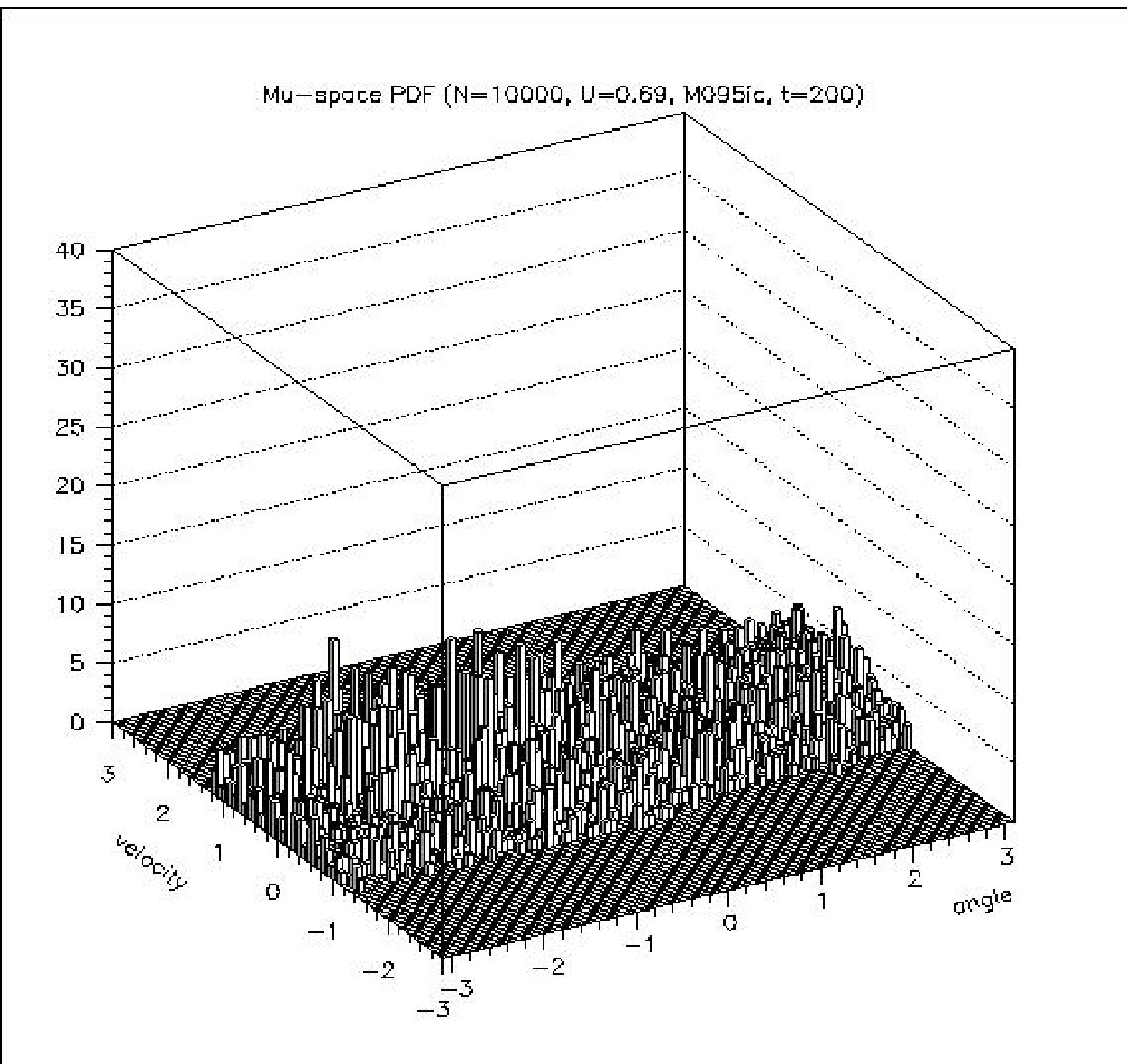}
}
\centerline{\includegraphics[width=7.cm,height=7. cm]
{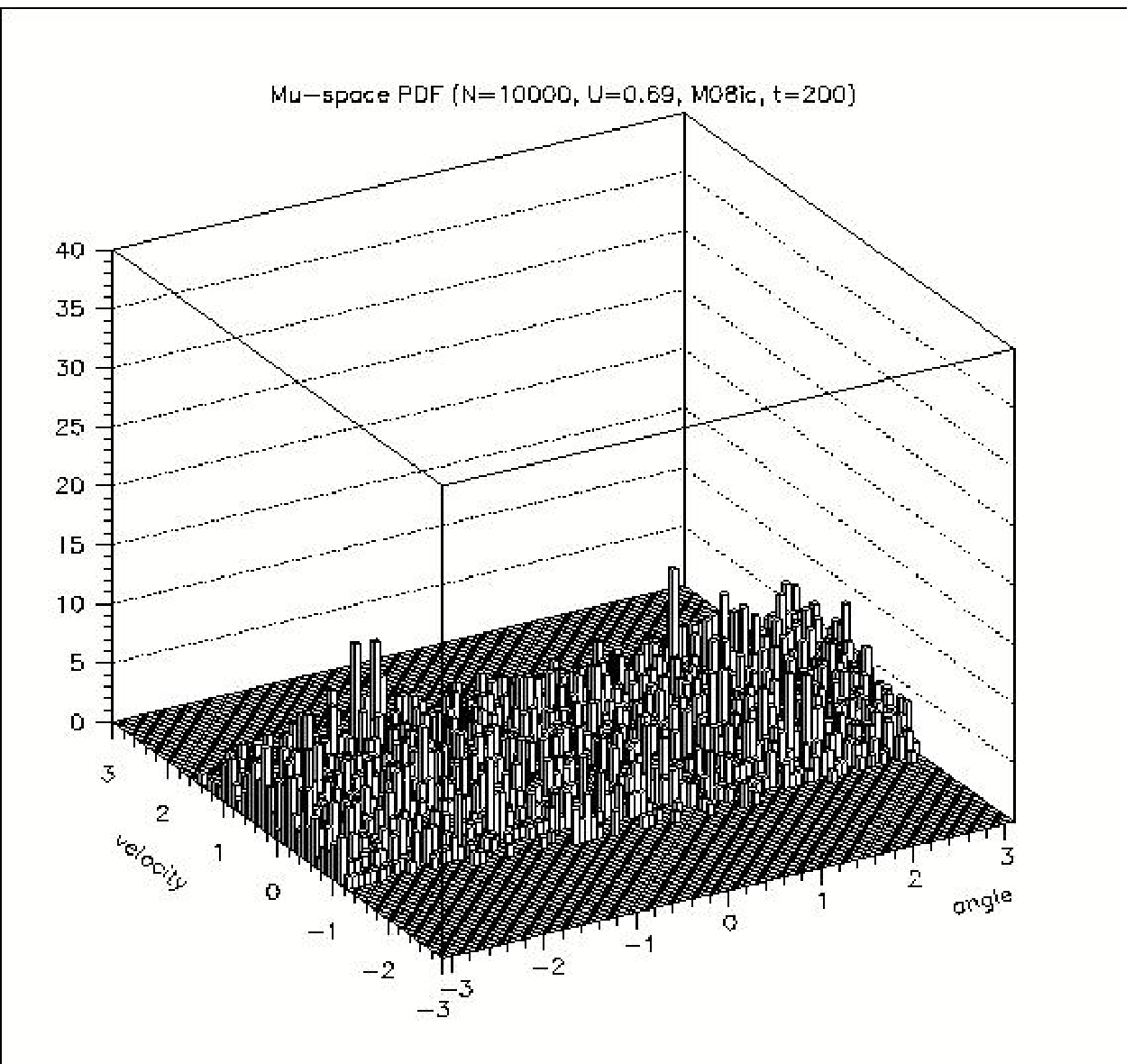}\includegraphics[width=7.cm,height=7. cm]{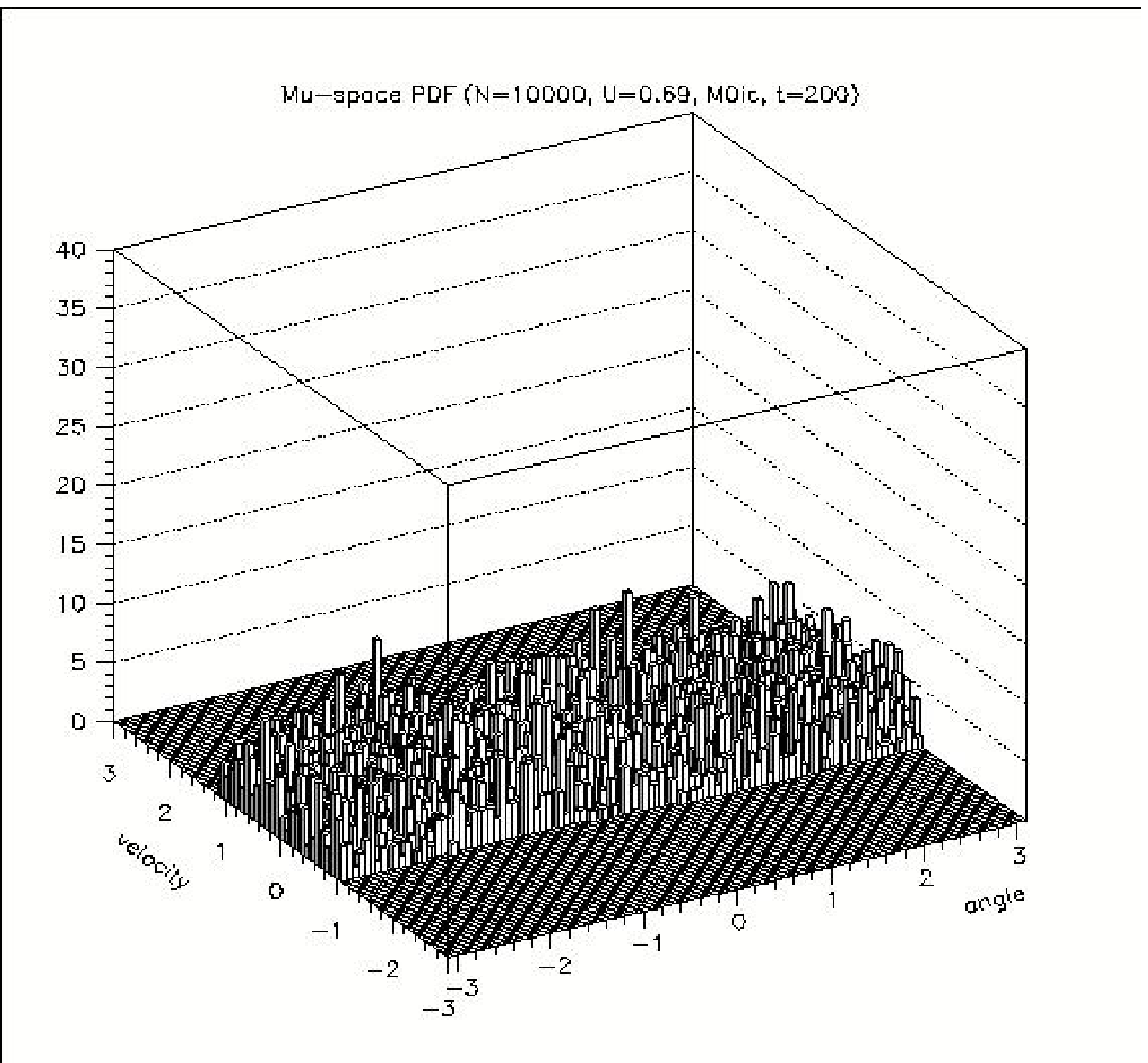}
}
\caption{For the energy density $U=0.69$, $N=10000$ and for different initial conditions we plot a
snapshot of the $\mu-space$ at time $t=200$ inside the plateau regime. The figure 
illustrates four different initial magnetization cases, 
M=1 (top-left),  M=0.95 (top-right), M=0.8 (bottom-left) and   M=0 (bottom right). 
Competing clusters of different sizes are clearly visible only for the  case
M=1. See text for further details.
\label{fig1}
}
\end{figure}

For the $M1 ic$ case, the dynamics in $\mu$-space can
be also clarified through the concept of "dynamical frustration": the
clusters appearing and disappearing on the unit circle compete one with each other
in trapping more and more particles, thus generating a dynamically frustrated
situation typical of  glassy systems \cite{plu2}.
In Fig.\ref{fig1} we show the dependence  
of the competing clusters  phenomenon as a function of the initial magnetization. 
The distribution function $f(\theta, p, t)$ is
considered for different out of equilibrium initial conditions (i.c.) inside the 
QSS regime. 
In particular for N=10000 and U=0.69 we plot a snapshot of 
the $\mu$-space configuration at $t=200$ for four initial 
magnetizations, namely M=1, M=0.95, M=0.8 and M=0, obtained by spreading
the initial angles distribution over a wider and wider portion of the unit circle. 
In this way we fix the initial potential
energy $V(\theta)$ and, in turn, the magnetization. Then, we 
assign the remaining part of the total energy as kinetic energy by
using a  water bag uniform distribution for the velocities.

We define each cluster as composed by
particles with both angles and velocities  in the same $\mu$-space cell.
A total of 100x100 cells for the $\mu$-space lattice has been considered.
In Fig.\ref{fig1} one clearly observes the presence of competing clusters only for
M=1 i.c..  At variance already for M=0.95 i.c. clusters are not very evident and the configuration tends to become more and more  uniform decreasing the initial magnetization.  
These simulations show that, although dynamical structures are present in the QSS regime
up to M=0.4 \cite{plu3}, frustration due to competing clusters  seems to depend 
much more strictly on the initial quenching  which is very violent for 
M=1 i.c..
\begin{figure} [t]
\centerline{\includegraphics[width=14cm,height=9 cm]
{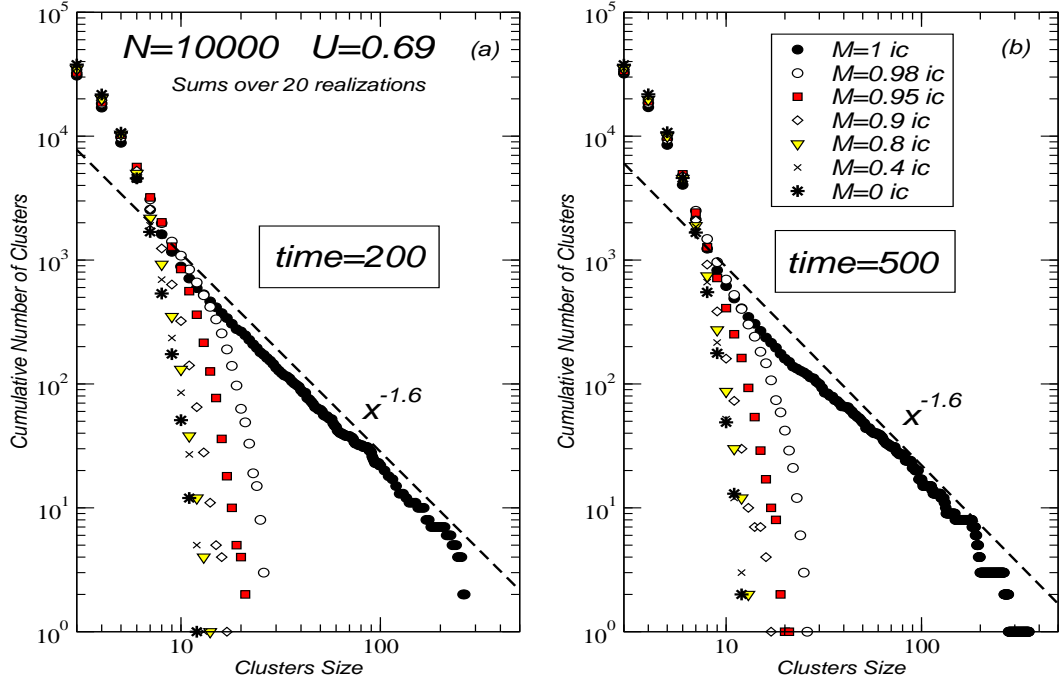}
}
\caption{We plot for the case $U=0.69$ and $N=10000$ 
 the cumulative distribution of the clusters size, calculated
for time $t=200$, panel (a),  and $t=500$, panel (b), in the QSS region. The points refers 
to different initial conditions: while the initial 
momenta were always uniformly ditributed, the initial magnetizations changed as reported in the legend. A power law can be observed only for initial magnetization M=1,  black circles.
A decaying power law with exponent $-1.6$ is also reported for comparison as dashed curve.  See text.
\label{fig2}
}
\end{figure}
In order to give further support to this result, in Fig. \ref{fig2}  
we show the behavior of the cluster size
cumulative distributions calculated in the case N=10000 and U=0.69 for
several initial magnetizations in the QSS regime at time t=200, 
panel (a), and t=500, panel (b).
For each one of the 100x100 $\mu$-space cells a sum over 20 different realizations 
of the dynamics (events) has been performed. 
Then, for each cluster size (greater than 5 particles) the sum of all the clusters with 
equal of bigger size has been calculated and plotted.
\\
As one can see from Fig.\ref{fig2}
a power law distribution, which indicates the presence of a hierarchical structure, 
can be observed only 
for initial conditions with $M=1$. In all the other cases studied, i.e. 
for  initial 
magnetization smaller or equal to $M=0.95$ the power law disappears. 
By comparing panel (a) with panel (b) one can also notice that, as expected, the 
distributions does not change
significatively in time in the plateaux region.
In the figure, we report also  a power law fit (drawn as a straight dashed
line above the data points) for the M=1 case: 
 for both cases at t=200 and at t=500, the cluster size distribution has an
exponent decay approximately equal to $-1.6$\footnote{Please nore that due to a misprint, the  exponent in the published version is erroneously reported as equal to $-1/6$.}.
\\
Such a strong dependence of the clusters hierarchical structure 
on the initial conditions  confirms that, 
although a metastable macroscopic regime is observed in all cases,
the microscopic structure of correlations is deeply different. 
The cluster size distribution 
observed for M=1 i.c. reminds closely that of percolation
at the critical point, where a lenght scale, or time scale,
diverges leaving the system in a self-similar state
\cite{binney}.
More in general, it has been suggested \cite{sotolongo} that,
optimizing Tsallis' entropy with natural 
constraints in a regime of long-range correlations, 
it is possible to derive a power-law hierarchical cluster size
distribution which can be considered as paradigmatic of physical
systems where multiscale interactions and
geometric (fractal) properties play a key role in the relaxation
behavior of the system.
The power-law scaling  strongly suggests a 
non-ergodic topology of the region  of  phase
space in which the system remains trapped when the QSS
regime is reached. 

These results  also supports the so called 
"weak ergodicity-breaking" scenario typical of glassy systems.
In ref.\cite{bouchaud} such a mechanism has been proposed in order
to explain the aging phenomenon, i.e. the dependence of the
relaxation time on the history of the system\cite{spin-glass}.
Aging has been found also in the HMF model for $M1$
i.c., more precisely in the autocorrelation functions decay for
both the angles and velocities \cite{celia1} and for velocities
only \cite{plu1}.  
More recently, inspired by the physical meaning of the Edwards-Anderson spin-glass order parameter \cite{EA_SK},
we proposed a new order parameter for the HMF
model, with the aim to measure the degree of freezing of the rotators in the QSS regime and 
to characterize in a quantitative way the emerging glassy-like dynamics  \cite{plu2}.
\begin{figure} [t]
\centerline{
\includegraphics[width=12cm,height=10 cm]{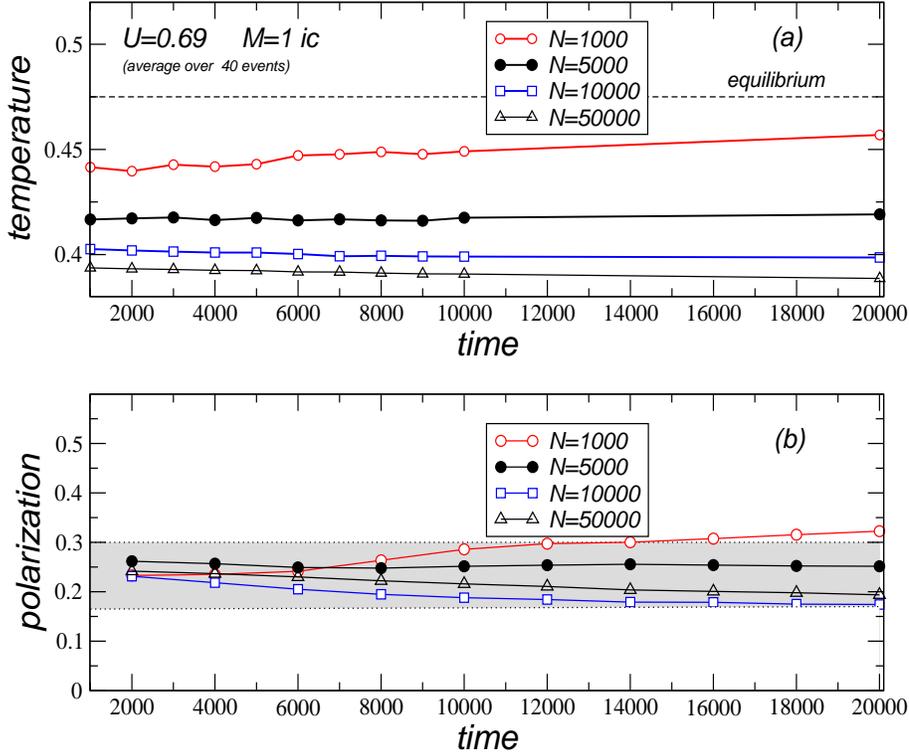}
}
\caption{For the energy density $U=0.69$ and different sizes (N=1000,5000,10000,50000), we plot the time evolution of the temperature for different initial magnetizations  in panel (a). In panel (b), we show the corresponding polarization values obtained for $\tau=2000$ and for increasing times along the QSS plateau.  The latter seems to be quite constant in time and with a value $0.24 \pm 0.05$. We plot a grey zone to illustrate these limits of variation. Notice that  for N=1000 the the temperature start to relax towards the equilibrium value around t=5000, so also the polarization tends to increase.  
\label{fig3}
}
\end{figure}
We   define the {\it elementary polarization} as the temporal
average, integrated over an opportune time interval $\tau$, of the
successive positions of each rotator: %%
\begin{equation}
\label{elpol}
<\stackrel{\vector(1,0){8}}{s_i}>={1\over{\tau}} \int_{t_0}^{t_0+\tau}
\stackrel{\vector(1,0){8}}{s_i}(t)dt~~~~~~i=1,...,N ~~,
\end{equation}
being $t_0$ an initial transient time.
Then we  average the module of the elementary polarization
over the N rotators, to  obtain the {\it polarization p}:
\begin{equation}
\label{pol}
{\it p}={1\over{N}} \sum_{i=1}^N  | <\stackrel{\vector(1,0){8}}{s_i}>|~~~~.
\end{equation}
It is important to  point  out that, in order to calculate 
 the elementary polarization of eq.\ref{elpol}, 
one has to subtract the global motion of the system, i.e. the phase of the average magnetization,  from the phase of each rotator.
\\
By means of several numerical simulations we showed in ref.\cite{plu2}
that in the QSS regime for $0.68 < U < 0.75$ and $M1 ic$, the 
polarization is constant and greater than zero - inside the error representing the fluctuations of the elementary polarization module -  while the magnetization vanishes with the size of the system.  Thus the polarization can be really considered 
a new order parameter  able characterize the glassyness of the QSS regime 
 \cite{plu2} .
\\

In Fig.\ref{fig3} we present new calculations for $U=0.69$ and $M=1$ initial conditions. In panel (a) we report the time evolution of temperature for different sizes of the system ($N=1000$, $N=5000$, $N=10000$ and $N=50000$) and after a transient $t_0=1000$.  
In panel (b) we plot the corresponding polarization calculated at different times and for a time interval 
$\tau=2000$. One can see that the polarization  remains almost constant around 
a value $p=0.24$ and inside an error of $\pm 0.05$ (grey area in the figure), it does not change substantially  inside  the QSS temperature plateaux and does not depend within the error on the size of the system. 
In general  the length of the plateaux regime depends on the size of the system,
thus the integration time $\tau=2000$ has been chosen in order to make a
meaningful calculation for all the sizes, included the smaller one ($N=1000$).
\begin{figure} [h]
\centerline{
\includegraphics[width=12cm,height=10 cm]{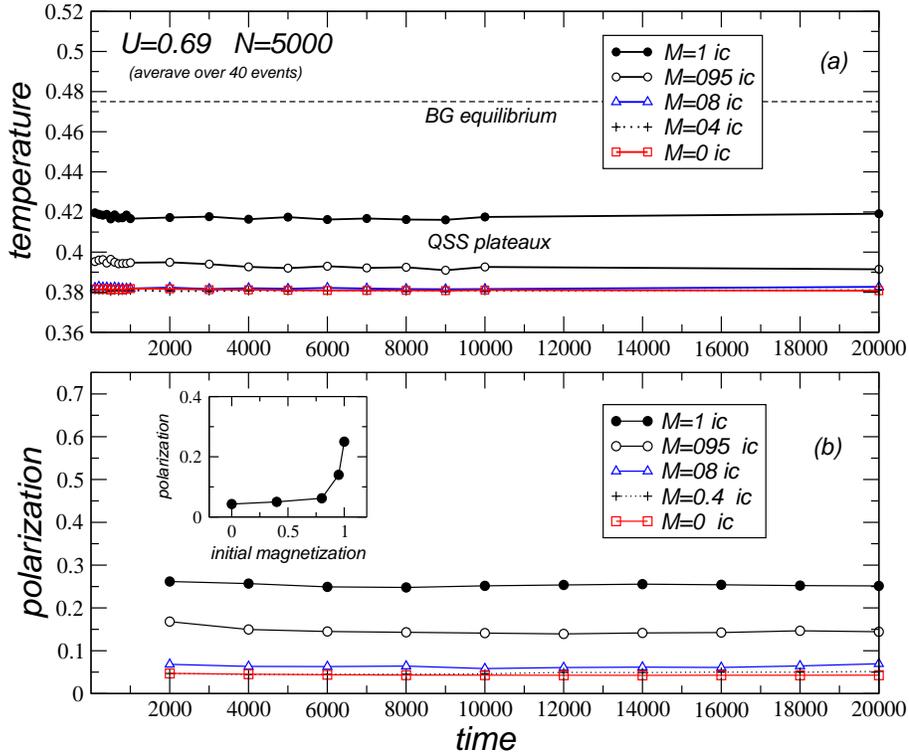}
}
\caption{For the energy density $U=0.69$ and N=5000, we plot the time evolution of the temperature for different initial conditions for the magnetization in panel (a). In panel (b), we show the corresponding polarization increasing the time interval of calculation. The polarization seems to depend strongly on the initial conditions and tends rapidly to zero as a function of the initial magnetization as reported in the inset. See text for further details.
\label{fig4}
}
\end{figure}

In Fig.\ref{fig4} we compare the calculation of the polarization  obtained for $U=0.69$, $N=5000$ and different initial magnetizations i.e.  $M=1,0.95, 0.8, 0.4, 0$. We report in panel (a) the temperature time evolution, while in panel (b) we plot the corresponding polarization as a function of time. Also in this case a time interval $\tau=2000$ has been considered to calculate the polarization. 
The plots are in agreement with the previous results about the $\mu$-space cluster distributions.
In fact also the polarization seems to be strongly dependent on the initial conditions 
and rapidly tends to zero by decreasing the initial magnetization, as reported in the inset of panel (b).
Again, the fast initial quenching we observe for the M=1 i.c. confirms its key role in
generating glassy behaviour in HMF model. 
On the other side, as we will show in the next subsection, nonextensive features are
quite more robust and seems to be less sensible to the initial magnetization.

\subsection{{Nonextensive thermodynamics and HMF model}}

\begin{figure} [h]
\centerline{
\includegraphics[width=12cm,height=10 cm]{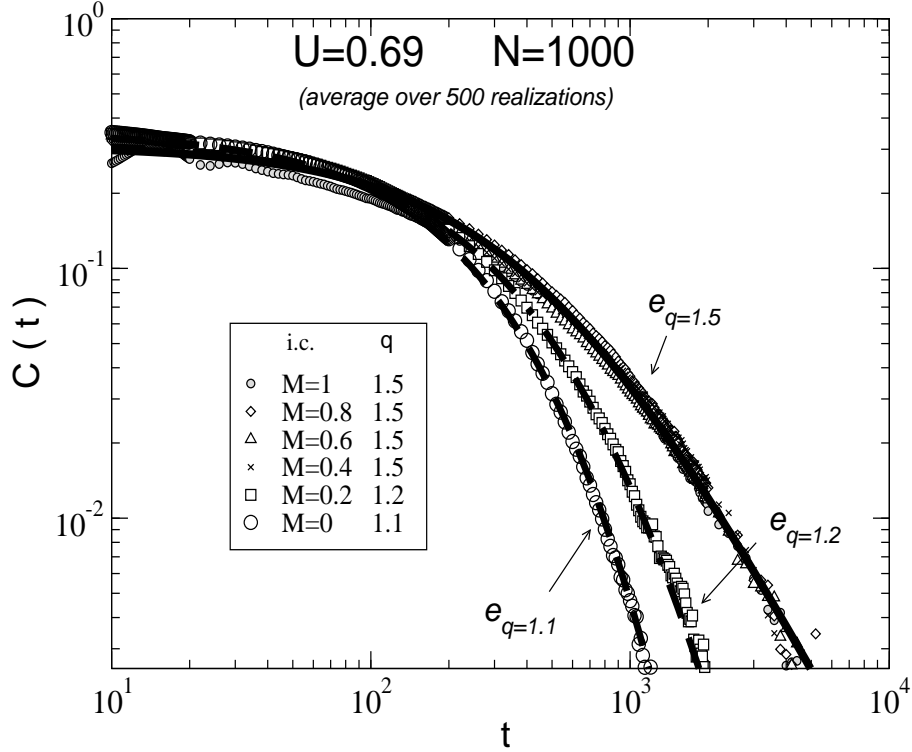}
}
\caption{For the energy density $U=0.69$ and N=1000, we plot the time evolution of the correlation function $C(t)$ for different initial  magnetization. 
A q-exponential fit is also reported for each curve 
 See text for further details.
\label{fig5}
}
\end{figure}

In previous works it was  shown that the majority of the dynamical
anomalies of the QSS regime, among which  velocity correlations in the $\mu$-space, 
are present not only for $M1$
initial conditions, but also when the initial magnetization
$M(t=0)$ is decreased in the range $(0,1]$ \cite{plu3}.
\\
The velocity correlations can be calculated by using the following
autocorrelation function\cite{plu3}
\begin{equation}
{C}(t)= \frac{1}{N} \sum_{j=1}^N {p_j(t) p_j(0)} ~~,
\label{nn_corr}
\end{equation}
where $p_j(t)$ is the velocity of the $j$-th particle at the time $t$.
In Fig.\ref{fig5}, 
we plot the velocity autocorrelation function (\ref{nn_corr}) for
$N=1000$, $U=0.69$ and $M(0)={1, 0.8, 0.6, 0.4, 0.2, 0}$.
%The initial fast relaxation has been truncated and 
An ensemble average over 500 different realizations was  performed.
For $ M(0)\ge 0.4$ the correlation functions are very similar,
while the decay is faster for $M(0)=0.2$ and even more for $M(0)=0$.
If we fit these relaxation functions by means of the 
Tsallis' q-exponential function
\begin{equation}
e_q(z)= {\left[  1+(1-q) z \right]} ^\frac{1}{(1-q)}~~,
\end{equation}
with $z=-\frac{t}{\tau}$, and where $\tau$ is a characteristic
time, we can quantitatively discriminate between the  different
initial conditions. In fact we get a q-exponential with $q=1.5$
for $M(0)\ge0.4$, while we get $q=1.2$ and  $q=1.1$ for $M(0)=0.2$
and for $M(0)=0$ respectively. Notice that for $q=1$ one recovers
the usual exponential decay \cite{tsallis1,hmf2,hmf3,plu1}. Thus for
$M(0)>0$ correlations exhibit a long-range nature and a slow
power-law  decay. This decay is  very similar for $M(0)\ge 0.4$, but  
diminishes progressively below $M(0)=0.4$ to become almost exponential for $M(0)=0$.
\\
Velocity correlations seem to be linked to anomalies in the diffusion.
In order to study diffusion, one  can consider  the mean square displacement
of phases $\sigma^{2}(t)$ defined as
\begin{equation}
\sigma^{2}(t) = \frac{1}{N} \sum_{j=1}^{N}
  [ \theta_{j}(t) - \theta_{j}(0) ]^{2}
  = < [ \theta_{j}(t) - \theta_{j}(0) ]^{2} > ~,
\label{msd2}
\end{equation}
where the symbol $<...>$ represents the average over all the $N$
rotators. The quantity $\sigma^{2}(t)$ typically scales as
$\sigma^{2}(t)\sim t^{\gamma}$. The diffusion is normal when
$\gamma=1$ (corresponding to the Einstein's law for Brownian
motion) and ballistic for $\gamma=2$ (corresponding to free
particles). For different values of $\gamma $ the diffusion 
is anomalous, in particular for $1<\gamma<2$ one has superdiffusion. 
We  notice that the quantity $\sigma^{2}(t)$
can be rewritten by using the velocity correlation function $C(t)$
as
\begin{equation}
    \sigma^{2}(t)
     = \int_{0}^{t} dt_{1} \int_{0}^{t} dt_{2}~
    < p_{j}(t_2)~p_{j}(t_1) > \\
     = 2 \int_{0}^{t} dt_1 \int_{0}^{t_1} dt_{2}~
    C(t_2)~~~,
\label{sigma_corr}
\end{equation}
where $C(t)$ is defined as in Eq.\ref{nn_corr}.
\\

Superdiffusion has been already observed in the HMF
model  for  M1 initial conditions\cite{hmf-dif}. Rercently  we have also
checked that, even decreasing the initial magnetization, the
system continues to show superdiffusion \cite{plu3}. In particular we obtained 
an ananoalous  diffusion  exponent that goes progressively from $\gamma=1.4-1.5$ for $0.4<M(0)<1$ to
 $\gamma= 1 $
for M0 ~\cite{erice}. No sensitive dependence on the size of the system 
has been orserved.
The slow decay and the superdiffusive behavior 
can be connected by
means of a conjecture based on a theoretical result found in 
ref.\cite{tsa-buk} by
Tsallis and Bukman. In fact in that paper the
authors show, on {\it general} grounds, that nonextensive
thermostatistics constitutes a theoretical framework within which
the {\it unification} of normal and {\it correlated} (driven)
anomalous diffusions can be achieved. They obtain, for a generic
linear force $F(x)$, the physically relevant {\it exact} (space,
time)-dependent solutions of a generalized Fokker-Planck 
equation.
Following ref.\cite{tsa-buk}, it is
possible to recover this relationship 
\begin{equation}
\gamma =\frac{2}{3-q}~,
\label{gqrel}
\end{equation} 
between the exponent
$\gamma$ of anomalous diffusion and the entropic index $q$.
Hence, being the space-time distributions
linked to the respective velocity correlations by the eq.
(\ref{sigma_corr}), one could think to insert in  eq. 
(\ref {gqrel}) the entropic index $q$,
characterizing the correlation decay  and the corresponding  anomalous diffusion
exponent. 
In order to check this conjecture, we have studied the ratio $\frac{\gamma}{2/(3-q)}$ vs the exponent 
$\gamma$ for various initial conditions ranging from
M(0)=1 to M(0)=0  and different sizes at $U=0.69$ . 
Within an uncertainty of $\pm 0.1$, the data \cite{erice} (not reported here for lack of space) show that this  ratio is always one, thus providing a  strong 
indication in favor of this conjecture. Although rigorous results are still missing the numerical evidence here discussed indicate that  Tsallis statistics and in particular its entropic index $q$ seems to be able to give a quantitative characterization of the correlations and dynamical anomalies  found in  the QSS regime of the HMF model. 
\\

\section{Conclusions}
We have briefly reviewed some of the anomalous features oberved in
the dynamics of the HMF model, a kind of minimal model
for the study of complex behavior in systems with long-range interactions.
We have also discussed how the anomalous behavior can be interpreted
within the nonextensive thermostatistics introduced by Tsallis,
and in the framework of the theory glassy systems.
These two frameworks  seems to have  several links which will be further 
explored in the future.

\section*{Acknowledgements}
We would like to thank the organizers and in particular S. Abe for the invitation to participate to this conference and the wonderful hospitality reserved to us in Kyoto.  This work is part of a long-term project  in collaboration also  with  V. Latora.

%\appendix
%\section{First Appendix} %Empty argument \section{} yields `Appendix'. 
%
%\section{Second Appendix}

\end{document}